\documentclass{article}

\usepackage{arxiv}

\usepackage[utf8]{inputenc} %
\usepackage[T1]{fontenc}    %
\usepackage{hyperref}       %
\usepackage{url}            %
\usepackage{booktabs}       %
\usepackage{amsfonts}       %
\usepackage{nicefrac}       %
\usepackage{microtype}      %
\usepackage{cleveref}       %
\usepackage{graphicx}
\usepackage{natbib}
\usepackage{doi}

\title{Parameterizable Acoustical Modeling and Auralization of Cultural Heritage Sites based on Photogrammetry}

\date{January 31, 2023}

\author{ \href{https://orcid.org/0000-0002-7904-3892}{\includegraphics[scale=0.06]{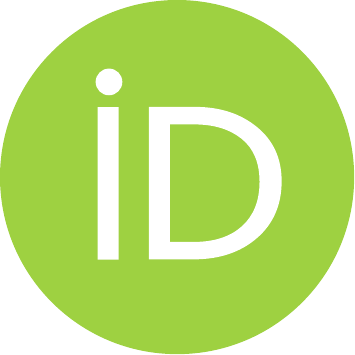}\hspace{1mm}Dominik Ukolov}\\
	Research Group DIGITAL ORGANOLOGY\\
	Leipzig University\\
	Augustusplatz 10, 04109 Leipzig (Germany)\\
	\texttt{dominik.ukolov@uni-leipzig.de} \\
}

\renewcommand{\headeright}{Preprint}
\renewcommand{\undertitle}{Preprint}

\hypersetup{
pdftitle={Parameterizable Acoustical Modeling and Auralization of Cultural Heritage Sites based on Photogrammetry},
pdfsubject={cs.SD, eess.AS},
pdfauthor={Dominik Ukolov},
pdfkeywords={acoustics, photogrammetry, cultural heritage, auralization, virtualization},
pdfpagelayout=OneColumn
}

\begin{document}
\maketitle

\begin{abstract}
	The photogrammetric and reconstructive modeling of cultural heritage sites is mostly focused on visually perceivable aspects, but if their intended purpose is the performance of cultural acts with a sonic emphasis, it is important to consider the preservation of their acoustical behaviour to make them audible in an authentic way. This applies in particular to sacral and concert environments as popular objects for photogrammetric models, which contain geometrical and textural information that can be used to locate and classify acoustically relevant surface properties. With the advancing conversion or destruction of historical acoustical spaces, it becomes even more important to preserve their unique sonic characters, while three-dimensional auralizations become widely applicable. The proposed study presents the current state of a new methodological approach to acoustical modeling using photogrammetric data and introduces a parameterizable pipeline that will be accessible as an open-source software with a graphical user interface.
\end{abstract}


\section{Introduction}
\begin{figure}
	\centering
	\includegraphics[width=\linewidth]{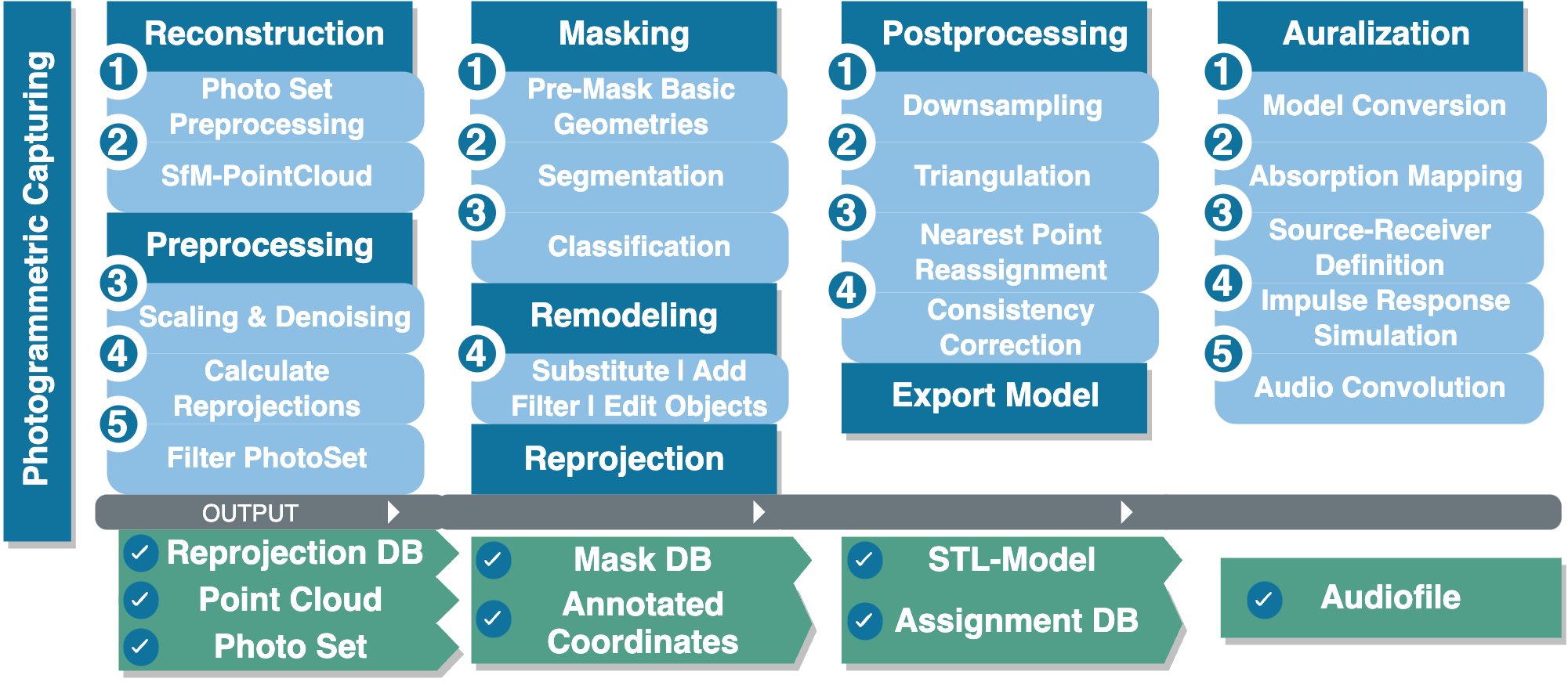}
	\caption{This pipeline contains multiple processing steps that can be modified by various parameters, while the procedurally generated output data can be further analyzed, edited, or used in other instances. The modular structure allows the optimization and addition of further steps, which will be evaluated and implemented during the MODAVIS project.}
	\label{fig:figure_01}
\end{figure}

The virtualization of musical instruments is an emerging field in digital organology, 
which has mostly been limited to capturing and synthesizing their sounds through sampling techniques or physical modeling. 
However, the MODAVIS project aims at researching a methodology for the multimodal and scientifically consistent digitization and virtualization of musical instruments as three-dimensional acoustical objects and the development of a new standard: the Virtual Acoustic Object \citep{Ukolov2023}. This research is aimed in particular at the pipe organ, the most complex and largest of all instruments, but also the most endangered one due to the consequences of climate change, 
military conflicts and economic crises. Beyond that, it is a challenge to fully understand the sound of this instrument without considering its surroundings, due to its acoustical coupling to historical and representative buildings, which must be thought of as resonant bodies and thus accurately modeled. Since the sound propagates three-dimensionally while being driven by the dimensions, materials and shapes inside the building, photogrammetric models seem very promising for the generation of interactive virtual acoustic objects and acoustical room models for the purpose of auralization. The proposed methodology to approach this problem is based on the principle of point projections between classified masks and their three-dimensional correlates with annotations of acoustical properties, performed by a parameterizable pipeline (see Figure \ref{fig:figure_01}).

\section{Related Work}
Related studies of the use of photogrammetric models for interior acoustical simulations were conducted by \citet{LlorcaBofi2022}, 
who applied geometric reductions based on \citet{Siltanen2008} and evaluated several modeling and material assignments using the RAVEN plugin \citep{Schroeder2011}, while the fundamentals of auralization were researched by \citet{Vorlaender2020}. With SoundSpaces 2.0 by \citet{Chen2022}, continuous renderings of impulse responses were performed using bidirectional path-tracing on mesh datasets and evaluated after acoustically measuring an object of the Replica dataset \citep{Straub2019}. However, the developed methodology described below differs from these approaches, as it is based on remodeling using point reprojections and multiple segmentations by neural networks; to the current knowledge, this is the first approach to generate acoustical models using these methods.

\section{Photogrammetric-Acoustical Modeling Toolkit}
\begin{figure}
	\centering
	\includegraphics[width=\linewidth]{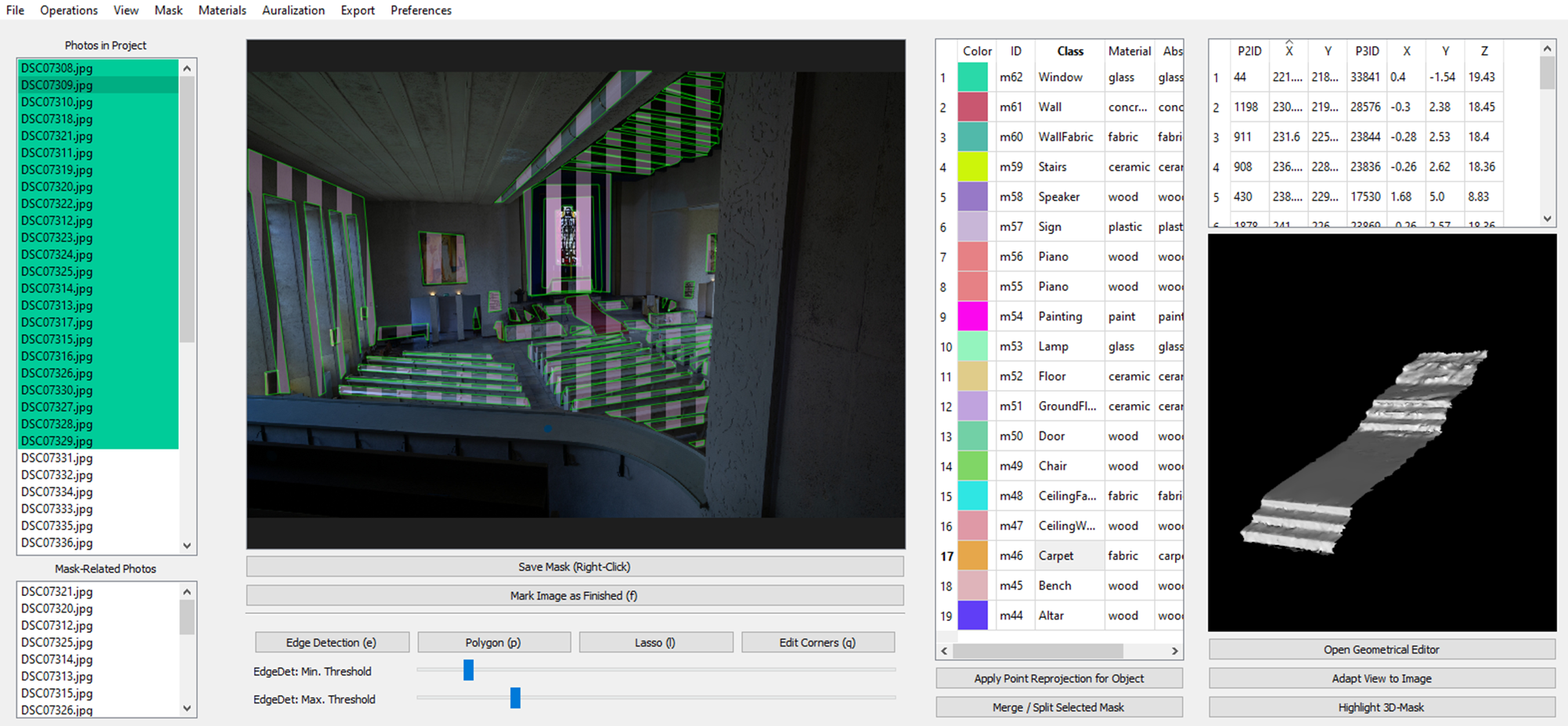}
	\caption{The main interface of the PAMT consists of the reduced set of photos that was used in the photogrammetric calculations (upper left), the masking window and its controls (middle left) and the list with the masked areas and classified objects, materials, and absorption coefficients (middle right). After a reprojection has been applied, the related images and the correlating local and global points can be investigated in the bottom left and in the upper right, which will be marked on the image and on the 3D viewer below. This viewer is interactive and shows the object for the selected and reprojected mask, while the ‘Geometrical Editor’ opens an external window for the object modification, plane segmentation and other geometrical operations. In this screenshot, the carpet was selected in the masks list and the corresponding reprojection was applied, resulting in rendering the associated point coordinates as a preview of the current masking progression.}
	\label{fig:figure_02}
\end{figure}

In order to parameterize the pipeline with real-time visual previews and to enable an easy operability of the acoustical modeling process with different levels of expertise, the Photogrammetric-Acoustical Modeling Toolkit (PAMT) was developed (see Figure \ref{fig:figure_02}), which allows manual and semi-automated processing steps.

\subsection{Preprocessing}
The PAMT uses COLMAP \citep{Schoenberger2016} to calculate the point cloud from a set of photos and to extract the camera parameters, 
which are used to generate a point reprojection database in the first step after creating a new project file, where all pipeline settings and individual modeling operations are saved. In the next step, the extracted data can be scaled based on markers, measurements, or model alignments, followed by filtering the photo set based on their overlaps to reduce their quantity for the following processes.

\subsection{Object Segmentation}
The reduced photo set is available in the GUI for manual or automated segmentation of objects, 
where manual segmentation is performed by polygonal masking with adjustable edge detections using the Canny algorithm \citep{Xu2017}. 
The automated segmentation uses Mask R-CNN \citep{He2017}, while its output above an adjustable threshold will be used for generating the masks. Once a mask has been set, an object class, an object, and a material can be assigned either manually or by classification suggestions; afterwards, several masks can be merged or divided, as all masks remain editable at any time. Once a photo has been processed, the masked points will be extrapolated to the other photos by projecting those 3D point identities that are near the reprojected 2D pixels inside the masks. This results in pre-generated local masks for global surfaces, significantly reducing the time for the manual tasks. The pre-trained model for the automated detection and segmentation can be selected, downloaded, and updated in the settings, for which MMLab \citep{Chen2019} is integrated into the toolkit to ensure the implementation of state-of-the-art models for this task.

\subsection{Material Classification and Absorption Coefficients}
The material can be defined after setting a mask, either manually or by following classifications from neural networks in two stages, 
currently based on the ResNet-18 \citep{He2016} architecture that has been trained on the MINC dataset \citep{Bell2015} for materials and the DTD dataset \citep{Cimpoi2014} for textural properties like fibrous, marbled, or porous. In order to assign frequency-dependent absorption coefficients to a textured material, a specialized database with 2573 entries \citep{PTB2018} was processed to be used for suggesting probably suitable measurements; this can be achieved through natural language processing (NLP) by calculating the cosine similarities of the input string and the acoustically measured material names based on their vector representations.

\subsection{Point Reprojection}
The basic method behind the reprojection is the assignment of local 2D coordinates to global 3D point identities, which are used to map local points within a photo. The polygonal masking defines a set of coordinates, from which the identities and their assigned 2D coordinates can be derived. As a result, the local coordinates of a photo set are always related to global coordinates in the PAMT and vice versa, for which database access structures have been developed which allow to return and process the target data by providing any geometrical input data.

\subsection{Object Modification}
Since objects such as rows of chairs often appear repetitively and not every of its appearances can be captured in detail, 
it is possible to substitute inaccurate object appearances in the point cloud by more detailed ones through pointing to the object or importing another point cloud or mesh. The correct placement is ensured by either estimating the point-relative positioning in the room or placing it manually, while object classes or individual objects can be filtered or added to the space, which also enables acoustical simulations of different room situations.

\subsection{Point Cloud and Surface Reconstruction}
\begin{figure}
	\centering
	\includegraphics[width=\linewidth]{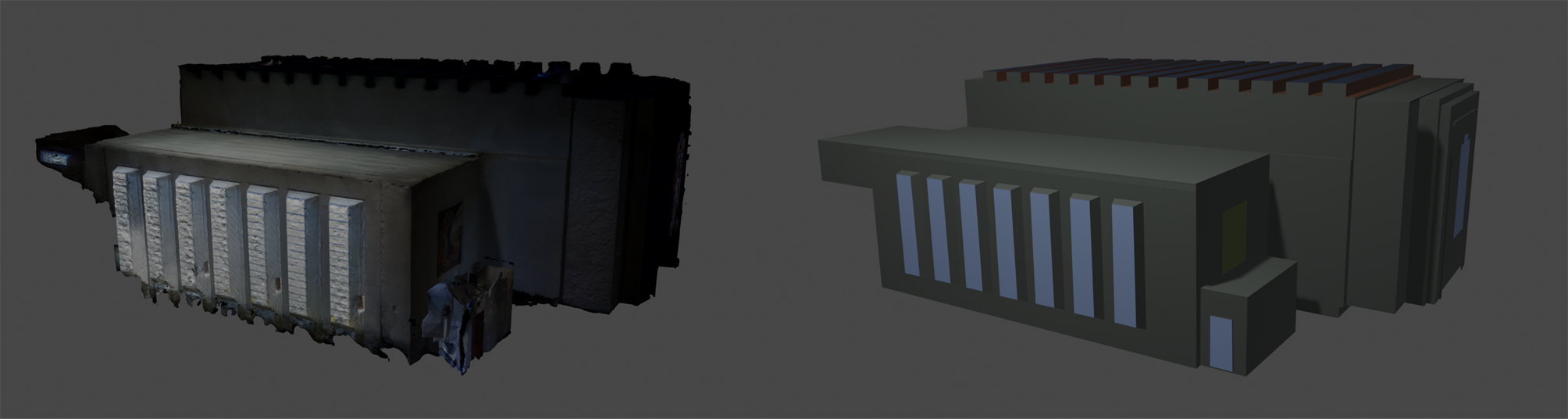}
	\caption{With the proposed pipeline, the photogrammetric data can be used to remodel its resulting point cloud into a geometrically reduced one (right), 
	which is suitable for acoustical simulations after consistency corrections. The photogrammetric model of the interior of a church (left), viewed from the outside, 
	was captured under poor lighting conditions with a low-quality camera and therefore contains inaccuracies in certain areas. 
	It is particularly noticeable that the plane segmentation as well as the boundary calculations lead to significant changes in shape (see left and bottom right side on both models), 
	which might lead to wanted results as in this case, but which must be further evaluated on variable and low- as well as high-quality datasets. 
	The materials that are visible on the right - glass (light blue), wood (brown), fabric (dark blue) and the painting (yellow) - 
	were assigned to surfaces that have been defined during the pipeline processing, while the exterior of the building was not captured and is not relevant in the current state of development.}
	\label{fig:figure_03}
\end{figure}

After defining the objects coordinates by masking them, an annotated point can be generated using the Blender API and rendered in the GUI with interactive controls, where object classes or individual masked areas can be shown. The denoising of the point cloud is performed using a score-based approach \citep{Luo2021}, while the rendering and parameterized processes are executed using Open3D \citep{Zhou2018}, consisting of downsampling and triangulation, currently using the ball-pivoting \citep{Bernardini1999} or Poisson \citep{Kazhdan2006} algorithm. The plane segmentation is fundamental for the model building (see Figure \ref{fig:figure_03}) and is performed using the RANSAC algorithm \citep{Fischler1981}, but as the point sets have been changed during these processes, the nearest points will finally be assigned to the corresponding identities. In the final step, the model will be exported to the STL format and validated to prevent inconsistencies like holes or inadequate volumes during auralizations. In addition, a referential database will be generated to point to the acoustical surface properties for every element in the STL file.

\subsection{Acoustical Simulation and Auralization Interfaces}
The implementation of the annotated model for the simulation of acoustical room situations from a specific listener position can be realized using an interface to pyroomacoustics \citep{Scheibler2018}, which converts the encoded geometric and acoustical data into compatible structures. After setting an arbitrary source and receiver position by defining their coordinates and emitting properties, the image source method \citep{Allen1979} and acoustical ray tracing \citep{Krokstad1968} can be applied to simulate impulse responses and convolute them with an input audio signal. By using the photogrammetric calculations, it is also possible to perform auralizations from the exact perspective that is captured in a photo, as the corresponding camera positions and orientations are known.

\section{Future Work}
It is planned to evaluate the PAMT performance using valid acoustical test datasets and to add important psychoacoustical and acoustical parameters such as scattering coefficients and directivity patterns to the auralization engine, which is generally considered as secondary at the current state of development. The configurations and methods of the geometrical processes shall generally be optimized to achieve the lowest possible error rate. Within the MODAVIS project, it is planned to create an object and material segmentation dataset for sacral environments and organ-specific parts, in addition, it shall be examined to what extent directivity patterns of different organ pipe types can be simulated and generalized via physical modeling. Furthermore, it shall be possible to use RGB-D data and non-photogrammetrically generated 3D models in the toolkit using virtual image renderings as well as to indicate the perspective on an orthographic map to prevent confusions with repetitive structures. It will be considered to publish a Blender module and an interface to Unity for annotation and auralization purposes at a later phase. After the evaluations and significant improvements are completed, the PAMT will be released on GitHub \citep{Ukolov2022}.

\bibliographystyle{abbrvnat}
\bibliography{references}

\end{document}